\documentclass[%
 reprint,
amsmath,
amssymb,
 aps,
 prb,
 floatfix]{revtex4-1}

\usepackage[dvips]{graphicx}
\usepackage{amsmath,amsfonts,euscript,amssymb}
\usepackage{epsfig}

\DeclareSymbolFont{bfitletters}{OML}{cmm}{bx}{it}
\DeclareSymbolFont{bfitoperators}   {OT1}{cmr} {m}{n}
\DeclareMathSymbol{\bfitomega}{\mathord}{bfitletters}{"21}
\DeclareMathSymbol{\bfitrho}{\mathord}{bfitletters}{"1A}
\DeclareMathSymbol{\bfitgamma}{\mathord}{bfitletters}{"0D}
\DeclareMathSymbol{\bfitchi}{\mathord}{bfitletters}{"1F}
\DeclareMathSymbol{\bfitxi}{\mathord}{bfitletters}{"18}

\newcommand{\be}{\begin{equation}}
\newcommand{\ee}{\end{equation}}
\newcommand{\bea}{\begin{eqnarray}}
\newcommand{\eea}{\end{eqnarray}}

\begin{document}

\title{Intrinsic Time in Geometrodynamics of Closed Manifolds}

\author{A.B. Arbuzov}
\affiliation{Bogoliubov Laboratory of Theoretical Physics, Joint Institute for Nuclear Research, Dubna,
  141980 Russia}
\altaffiliation[Also at ]{ Department of Higher Mathematics, Dubna State University, Dubna, 141980, Russia.\\ arbuzov@theor.jinr.ru}
\author{A.E. Pavlov}
\affiliation{Institute of Mechanics and Energetics,
  Russian State Agrarian University --- Moscow Timiryazev Agricultural Academy,
  Moscow, 127550, Russia}
\altaffiliation{alexpavlov60@mail.ru}  

\date{\today}% It is always \today, today,

\begin{abstract}
The global time in Geometrodynamics is defined in a covariant under diffeomorphisms form.
An arbitrary static background metric is taken in the tangent space.
The global intrinsic time is identified with the mean value of the logarithm of the square
root of the ratio of the metric determinants. The procedures
of the Hamiltonian reduction and deparametrization of dynamical
systems are implemented. The explored Hamiltonian system appeared to be non-conservative.
The Hamiltonian equations of motion of gravitational field in the global time are written.
Relations between different time intervals (coordinate, intrinsic, proper) are obtained.

\end{abstract}


\maketitle

%%%%%%%%%%%%%%%%%%%%%%%%%%%%%%%%%%%%%%%%%%%%%%%%%%%%%%%%%%%%%%%%%%%%%%%%%%%%%%%%%%%%%%%%%%%%%%%%%%%%%%%%%%%%%%%%%%
\section{Introduction}
%%%%%%%%%%%%%%%%%%%%%%%%%%%%%%%%%%%%%%%%%%%%%%%%%%%%%%%%%%%%%%%%%%%%%%%%%%%%%%%%%%%%%%%%%%%%%%%%%%%%%%%%%%%%%%%%%%

Geometrodynamics is a theory describing space and time in its inner essence and the spatial
metric carries information about the intrinsic time. ``Time has a double face, as the Time-god,
Janus, has in the ancient representations. One quantity has two
interpretations -- chronometrical, and dynamical''~\cite{KucharJanus}.
The Hamiltonian dynamics of gravitational field is commonly formulated by construction
in redundant variables in the extended functional phase space as a consequence of covariant
description of Einstein's theory. A time parameter should be conjugated to the Hamiltonian
constraint. The Hamiltonian formulation of the theory makes it possible to reveal
the physical meaning of the geometrical variables. The problem of the global time definition
is a basic one in General Relativity~\cite{Kuchar,Isham:1992ms}.
It was demonstrated that to construct a quantum theory one needs to use the so-called
bubble-time derivatives instead of the ordinary ones~\cite{Teitelboim:1975sop,KBubble}, that complicates
the physical problem. The quantum Wheeler--DeWitt (or Schr$\ddot{\rm o}$dinger) equation for
wave functions is then constructed in functional derivatives.
The conservative energy is well defined in asymptotically flat spaces because of
the existence of the preferable rectangular coordinate frame with a coordinate
time~\cite{Arnowitt}.
The quantities are defined by boundary integrals~\cite{Regge}.

The Hamiltonian formalism was developed with the improved Hamiltonian~\cite{Regge}
or through modification the Poisson brackets~\cite{Soloviev}.
In the case of closed manifolds, the constant mean curvature slicing is preferable.
And in this case one can't introduce an outer observer with his watch.
The global time then can be found from the Hamiltonian constraint in homogeneous models,
see, {\it e.g.}, refs.~\cite{Kuchar,PIntern,Misner:1969hg}.
The conformal Dirac's mapping~\cite{Dirac:1958jc} allows to extract a local intrinsic time.
The Cauchy problem was solved in conformal variables successfully~\cite{OMurchadha:1974mtn,JamesII}.
Misner introduced the intrinsic time as a logarithm of hypersurface volume~\cite{Misner:1969hg}
in a homogeneous mixmaster universe.

In this paper we describe in detail a self-consistent procedure that allows to introduce
an intrinsic time in a closed manifold and discuss the corresponding implications in the
Hamiltonian reduction. The notation and preliminaries are given in
Sect.~\ref{Sect:ADM}. Introduction of a many-fingered intrinsic time is discussed in
Sect.~\ref{Sect:mftime}. Sect.~\ref{Sect:globaltime} presents the main result of the paper
which consists in the definition of the global intrinsic time. Using the defined global
time, the Hamiltonian reduction and deparametrization are performed in Sect.~\ref{Sect:reduction}.
Residue dynamical variables in a perturbed Friedmann universe are discussed
in Sect.~\ref{Sect:residue}. Hamiltonian equations of motion are derived in
Sect.~\ref{Sect:eq}. Relations between the intrinsic and coordinate time intervals
are given in Sect.~\ref{Sect:times}. Sect.~\ref{Sect:concl} contains discussion and conclusions.
Some detail of calculations are put in the Appendix.

%%%%%%%%%%%%%%%%%%%%%%%%%%%%%%%%%%%%%%%%%%%%%%%%%%%%%%%%%%%%%%%%%%%%%%%%%%%%%%%%%%%%%%%%%%%%%%%%%
\section{ADM variational functional. Notations} \label{Sect:ADM}
%%%%%%%%%%%%%%%%%%%%%%%%%%%%%%%%%%%%%%%%%%%%%%%%%%%%%%%%%%%%%%%%%%%%%%%%%%%%%%%%%%%%%%%%%%%%%%%%%

The spacetime ${\mathcal M}={\mathbb{R}}^1\times\Sigma_t$ with the metric tensor field
\begin{equation} \nonumber
{\bf g}:=g_{\mu\nu}(t, {\bf x}){\rm\bf d}x^\mu\otimes {\rm\bf d}x^\nu
\end{equation}
can be foliated into a family of space-like hypersurfaces $\Sigma_t,$ labeled
by the time coordinate $t$ with just three spatial coordinates on each
slice $(x^1, x^2, x^3)$. Then, the evolution of space in time can be described
in a natural way.
The components of the metric tensor in the Arnowitt-Deser-Misner (ADM) form are
\be\label{gmunu}
(g_{\mu\nu})= \left( \begin{array}{cc}
-N^2+N_iN^i& N_i \\
N_j& \gamma_{ij}
\end{array}
\right).
\ee
Here and below Latin indices run as $i,j=1,2,3$.
The scalar field $N$ and the 3-vector field ${\bf N}$ extend
the coordinate system out of $\Sigma_t$. The first quadratic form
\be\label{spacemetric}
{\bfitgamma}:=\gamma_{ik}(t, {\bf x}){\rm\bf d}x^i\otimes {\rm\bf d}x^k
\ee
defines the induced metric on every slice $\Sigma_t$.

The components of the extrinsic curvature tensor $K_{ij}$ of every slice are
constructed out of the second quadratic form of the hypersurface and can be defined as
\be\label{Kij}
K_{ij}:=-\frac{1}{2}{\pounds}_{\bf n}\gamma_{ij},
\ee
where ${\pounds}_{\bf n}$ denotes the Lie derivative along the time-like unit
normal to the slice vector ${\bf n}$.
The components of the extrinsic curvature tensor can be found by using the Lie
derivative along the vector field ${\bf N}$:
\bea \label{Kijdef}
K_{ij} &=& \frac{1}{2N}\left({\pounds}_{\bf N}\gamma_{ij}
-\frac{{\rm d}\gamma_{ij}}{{\rm d} t}\right)
\nonumber \\
&=& \frac{1}{2N}\left(\nabla_i N_j+\nabla_j N_i-\frac{{\rm d}\gamma_{ij}}{{\rm d} t}\right),
\eea
where $\nabla_k$ is the Levi--Civita connection associated with metric $\gamma_{ij}$:
$\nabla_k \gamma_{ij}=0$.
The phase space $\Gamma$ is coordinatized by the 3-metric~(\ref{spacemetric})
and its conjugate momentum density of weight~1 is defined by the components
\be\label{piijK}
\pi^{ij} := -\sqrt{\gamma}(K^{ij}-K\gamma^{ij}).
\ee
Here we introduced the notation
\bea \label{gammadet}
&& K^{ij}:=\gamma^{ik}\gamma^{jl}K_{kl}, \qquad K:=\gamma^{ij}K_{ij},
\nonumber \\
&& \gamma:=\det||\gamma_{ij}||, \qquad \gamma_{ij}\gamma^{jk}=\delta_i^k.
\eea
By contracting the tensors on both sides in~(\ref{Kij}), one obtains
\begin{eqnarray} \nonumber
\sqrt\gamma K=-{\pounds}_{\bf n}\sqrt\gamma.
\end{eqnarray}
The extrinsic curvature tensor $K_{ij}$ provides a measure of the bending
of the hypersurface $\Sigma_t$ with respect to the external space.
The tensor is proportional to the time derivative of $\gamma_{ij}$.
So it is naturally connected with physical quantities which are the momentum
densities $\pi^{ij}$ of the metric $\gamma_{ij}$.

To construct a functional phase space $\Gamma$ in the Banach space, one should first
define functionals. The Poisson bracket is a bilinear operation on
two arbitrary functionals $F[\gamma_{ij}, \pi^{ij}],$ $G[\gamma_{ij}, \pi^{ij}]$
over the hypersurface $\Sigma_t$
\bea \label{PB}
&& \{F, G\}:=
\int\limits_{\Sigma_t}\, d^3x\biggl(\frac{\delta F}{\delta \gamma_{ij}(t,{\bf x})}
\frac{\delta G}{\delta \pi^{ij}(t,{\bf x})}
\nonumber \\ && \qquad
 - \frac{\delta G}{\delta \gamma_{ij}(t,{\bf x})}
\frac{\delta F}{\delta \pi^{ij}(t,{\bf x})}\biggr).
\eea
The canonical variables satisfy the fundamental relations
\be\label{gammapi}
\{\gamma_{ij}(t,{\bf x}),\pi^{kl}(t,{\bf x}')\}=\delta_{ij}^{kl}\delta ({\bf x}-{\bf x}'),
\ee
where
\be\nonumber
\delta_{ij}^{kl}:=\frac{1}{2}\left(\delta_i^k\delta_j^l+\delta_i^l\delta_j^k\right),
\ee
and $\delta ({\bf x}-{\bf x}')$ is the Dirac's $\delta$-function~\cite{T} for
the volume of $\Sigma_t$ defined without recourse to the metric of space by
\be\nonumber
\int\limits_{\Sigma_t}{\rm d}^3x\,\delta ({\bf x}-{\bf x}')f({\bf x}')=f({\bf x})
\ee
for an arbitrary scalar probe function.

The super-Hamiltonian of the gravitational field is the functional
\be\label{superHam}
H_{\mathrm{ADM}}:=\int\limits_{\Sigma_t}\, {\rm d}^3x
\left( N(x){\cal H}_{\bot}(x)+N^i(x){\cal H}_i(x)\right),
\ee
where $N$ (lapse) and $N^i$ (shift) are Lagrange multipliers;
${\cal H}_{\bot}$ and ${\cal H}_i$ have the sense of constraints. In particular,
\bea \label{constraintHam}
&& {\cal H}_{\bot}(x) :=\frac{1}{2}\pi^{ij}(x)\frac{G_{ijkl}(x)}{\sqrt{\gamma}(x)}\pi^{kl}(x)
\nonumber \\ && \qquad
- \sqrt{\gamma}(x)R [\gamma](x)+\sqrt\gamma (x)T_{\bot\bot}
\eea
is obtained from the scalar Gauss relation of the embedding hypersurfaces theory and
called the Hamiltonian constraint.
Here $R[\gamma]$ is the Ricci scalar of the space,
\be\nonumber
 G_{ijkl}:=(\gamma_{ik}\gamma_{jl}+\gamma_{il}\gamma_{jk}-\gamma_{ij}\gamma_{kl})
\ee
is the supermetric of the 6-dimensional hyperbolic Wheeler--DeWitt (WDW) superspace.
The matter density
\be\nonumber
T_{\bot\bot}:=n^\mu n^\nu T_{\mu\nu}
\ee
is defined in a normal reference frame.
Momentum constraints
\be\label{constraintMom}
{\cal H}^i (x):= -2\nabla_j\pi^{ij}(x)+\sqrt\gamma (x)(T_\bot)^i
\ee
are obtained from the contracted Codazzi equations of the embedding hypersurfaces theory.
The components of the matter momentum density
\be\nonumber
(T_\bot)_i:=n^\mu T_{i\mu}
\ee
are defined in an observer's normal frame reference.
The constraints impose restrictions on possible data $\gamma_{ij}({\bf x},t)$,
$\pi^{ij}({\bf x},t)$ on a space-like hypersurface $\Sigma_t$.
Momentum constraints~(\ref{constraintMom}) generate 3-diffeomorphisms.
To prove the statement let them be smeared with a vector field $\bfitxi$
\be\nonumber
(\xi_i|{\cal H}^i)=\int\limits_{\Sigma_t}\,{\rm d}^3x\xi_i(x){\cal H}^i(x)
\ee
and take the Poisson bracket with the metric
\be\label{Killinggamma}
\{\gamma_{ij}(x),(\xi_i|{\cal H}^i)\}=\nabla_i\xi_j+\nabla_j\xi_i\equiv
{\pounds}_{\bfitxi}\gamma_{ij}.
\ee
By this rule an infinitesimal diffeomorphism acts on a metric field.
The Killing equations follow from~(\ref{Killinggamma}) as condition of
the Lie derivative of the metric along $\bfitxi$ to be zero.
Constraints (\ref{constraintHam}) and (\ref{constraintMom}) are of the first class
since they belong to a closed algebra \cite{TeiHen}.

Then the Hamiltonian dynamics is built of the ADM variational functional
\be\label{ADMvar}
S=\int\limits_{t_I}^{t_0}\,
{\rm d}t\int\limits_{\Sigma_t}{\rm d}^3x \left(\pi^{ij}\frac{{\rm d}\gamma_{ij}}{{\rm d}t}
- N{\cal H}_{\bot} - N^i{\cal H}_i\right),
\ee
where the common ADM units $c=1$, $16\pi G=1$ were used.
Action~(\ref{ADMvar}) is obtained from the Hilbert functional after the procedure
of $(3 + 1)$ foliation and the Legendre transformation are executed.
Let ${\rm Riem}(M)$ is a space of smooth Riemannian metrics. The WDW superspace is a factor-space ${\rm WDW}={\rm Riem}(M)/{\rm Diff}(M)$,
where ${\rm Diff}(M)$ is the group of smooth diffeomorphisms of $M$. It is the analog of the Teichm$\rm\ddot{u}$ller space for 2-manifolds.

%%%%%%%%%%%%%%%%%%%%%%%%%%%%%%%%%%%%%%%%%%%%%%%%%%%%%%%%%%%%%%%%%%%%%%%%%%%%%%%%%%%%%%%
\section{Many-fingered intrinsic time} \label{Sect:mftime}
%%%%%%%%%%%%%%%%%%%%%%%%%%%%%%%%%%%%%%%%%%%%%%%%%%%%%%%%%%%%%%%%%%%%%%%%%%%%%%%%%%%%%%%

The Dirac's mapping to conformal variables~\cite{Dirac:1958jc} has a restricted domain
of applicability: they can be used only for coordinate systems with a dimensionless
determinant of the metric. But often one needs to perform calculations, {\it e.g.}, in a spherical
coordinate system $(r, \theta, \varphi)$. Then one has to use the spherical
Schwarzschild's coordinates~\cite{Schwarzschild:1916uq}
\be\nonumber
x^1=\frac{r^3}{3},\qquad x^2=-\cos\theta,\qquad x^3=\varphi
\ee
with a unit metric determinant. To expand the region of applicability of the Dirac's
mapping, let us implement the following conformal transformation
\begin{equation}\label{Psifactor}
\gamma_{ij}:=\phi^4\tilde\gamma_{ij},\qquad \phi^4:=\sqrt[3]{\frac{\gamma}{f}},
\end{equation}
where, in addition to the determinant $\gamma$, the background static metric determinant
$f$ has appeared~\cite{Pavlov:2017auw}
\be \nonumber
f:={\rm det} (f_{ij}).
\ee
Introduction of the background metric is the key point of this paper.
It will allow us to consider not only asymptotically flat spaces but also
general closed manifolds. In addition to the space metric~(\ref{spacemetric}),
the background metric of the tangent space, Lie-dragged along the coordinate time
evolution vector, can be introduced:
\be\label{spacemetricback}
{\bf f}:=f_{ik}({\bf x}){\rm\bf d}x^i\otimes {\rm\bf d}x^k.
\ee
The Minkowskian metric as the background one was used in~\cite{Rosen,Gor} for description
of gravitational problems in asymptotically flat spacetimes.
We claim that the background metric has to be chosen from the physical point of view.
The mapping of the Riemannian space with metric $\bfitgamma$~(\ref{spacemetric})
to the background space with the metric ${\bf f}$~(\ref{spacemetricback}) should be bijective.
For this reason we suggest to consider a closed manifold.
The conformal metric~(\ref{Psifactor})
\be\label{gammatilde}
{\tilde\bfitgamma}:=\tilde\gamma_{ik}(t, {\bf x}){\rm\bf d}x^i\otimes {\rm\bf d}x^k
\ee
is a tensor field, {\it i.e.}, it is transformed according to the tensor representation
of the group of diffeomorphisms. The scaling factor $(\gamma/f)$ is a scalar field,
{\it i.e.}, it is invariant under the diffeomorphisms.
To the conformal variables
\begin{equation}\label{generalized}
{{\tilde\gamma_{ij}:=\frac{\gamma_{ij}}{\sqrt[3]{\gamma /f}},\qquad
\tilde\pi^{ij}:=\sqrt[3]{\frac{\gamma}{f}}\left(\pi^{ij}-\frac{1}{3}\pi\gamma^{ij}\right)}},
\end{equation}
we add the canonical pair: the local intrinsic time $D$ and the trace of momentum
density $\pi$:
\begin{equation}\label{DiracTpi}
D:=\frac{2}{3}\ln\sqrt{\frac{\gamma}{f}},\qquad \pi=2K\sqrt\gamma.
\end{equation}
Formulae~(\ref{generalized}) and~(\ref{DiracTpi}) define the scaled Dirac's
mapping~\cite{Pavlov:2017auw}
\begin{equation}\label{generalizedD}
(\gamma_{ij}, \pi^{ij})\mapsto (D,\pi; \tilde\gamma_{ij}, \tilde\pi^{ij}).
\end{equation}
The symplectic potential takes the form
\be\label{symplectic}
\omega^1:=\pi^{ij}{\rm d}\gamma_{ij}=\tilde\pi^{ij}{\rm d}\tilde\gamma_{ij}+\pi{\rm d}D.
\ee
The Lie algebra of new variables in the extended phase space $\Gamma_D$ are the same as
in~\cite{Regge} because of the static nature of the background metric
\begin{eqnarray}
&&\!\!\! \{D(t,{\bf x}), \pi(t,{\bf x}')\}=\delta ({\bf x}-{\bf x}'),
\label{PoissonD} \\
&&\!\!\! \{\tilde\gamma_{ij}(t,{\bf x}),\tilde\pi^{kl}(t,{\bf x}')\}=\tilde\delta_{ij}^{kl}\delta ({\bf x}-{\bf x}'),
\label{PoissonG} \\
&&\!\!\! \{\tilde\pi^{ij}(t,{\bf x}),\tilde\pi^{kl}(t,{\bf x}')\}=\frac{\tilde\gamma^{kl}\tilde\pi^{ij}
-\tilde\gamma^{ij}\tilde\pi^{kl}}{3} \delta ({\bf x}-{\bf x}'), \label{PoissonPi}
\end{eqnarray}
where
\be \nonumber
\tilde\delta_{ij}^{kl}:=\delta_i^k\delta_j^l+\delta_i^l\delta_j^k
-\frac{1}{3}\tilde\gamma^{kl}\tilde\gamma_{ij}
\ee
is the conformal Kronecker symbol. The sub-algebra of the canonical pair $(D,\pi)$
is split out of the algebra (\ref{PoissonG})--(\ref{PoissonPi}).

The Hamiltonian constraint in the new variables yields the quasilinear Lichnerowicz--York
differential equation to the field $\phi$
\be\label{H}
\left(\tilde\Delta-\frac{1}{8}\tilde{R}\right)\phi+
\frac{1}{8}\tilde{\pi}_{ij}\tilde{\pi}^{ij}\phi^{-7}-\frac{1}{12}K^2\phi^5
+\frac{1}{8}\tilde{T}_{\bot\bot}\phi^5=0.
\ee
Here $\tilde\nabla_k$ is the conformal connection associated with the conformal
metric $\tilde\gamma_{ij}$; quantity $\tilde{R}$ is the conformal Ricci scalar
related to the standard Ricci scalar $R$:
\be\label{Rscalar}
R= \frac{1}{\phi^4}\tilde{R}-\frac{8}{\phi^5}\tilde\Delta\phi.
\ee
The matter density is transformed according to
\be\label{Ttilde}
\tilde{T}_{\bot\bot}:=\phi^8 T_{\bot\bot},
\ee
where $T_{\bot\bot}$ is a component of the energy-momentum tensor along the future
pointing normal to $\Sigma_t$.

%%%%%%%%%%%%%%%%%%%%%%%%%%%%%%%%%%%%%%%%%%%%%%%%%%%%%%%%%%%%%%%%%%%%%%%%%%%%%%%%%%%%%%%%%%%%%%%%%
\section{Intrinsic global time} \label{Sect:globaltime}
%%%%%%%%%%%%%%%%%%%%%%%%%%%%%%%%%%%%%%%%%%%%%%%%%%%%%%%%%%%%%%%%%%%%%%%%%%%%%%%%%%%%%%%%%%%%%%%%%

The momentum density $\pi$ enters into the conformal Hamiltonian constraint
quadratically~(\ref{H}) as usual for relativistic theories. So, with the plus sign,
it can be expressed from the Hamiltonian constraint.
Taking into account the role of the Hamiltonian $H(x)$ as the integral over
the hypersurface of the $\pi(x)$ expressed from the Hamiltonian constraint
one should obtain the global time $T$.
Let us extract the zero mode out of the scalar field $D(x)$
\be\label{extract}
D(x)=<D>(t)+\overline{D}(x),
\ee
where the mean value of it over the hypersurface $\Sigma_t$ is
\be\label{mean}
<D>:=
\frac{\int_{\Sigma_t}\,{\rm d}^3 y\sqrt{\gamma} (y)D(y)}
{\int_{\Sigma_t}\,{\rm d}^3 y\sqrt{\gamma}(y)}.
\ee
According to the construction (\ref{mean}), the second term in~(\ref{extract}) is
the residue of the field $D(x)$ with a zero mean value over
a hypersurface $\Sigma_t$
\be\label{residuecondition}
\int\limits_{\Sigma_t}\,{\rm d}^3 y\sqrt{\gamma} (y)
\overline{D}(y)=0.
\ee
Now we can define the global time as discussed in our preprint~\cite{Arbuzov:2017col}
\be\label{global}
T(t):=-<D>(t)=-\frac{2}{3}<\ln\sqrt{\frac{\gamma}{f}}>
\ee
as the mean value over a hypersurface $\Sigma_t$ of the logarithmic function at every instant $t$.
The commutator of the global time $T(t)$ with the integral characteristics $P(t)$ of the field $\pi(x)$
\be\label{Pi}
P (t):=\int\limits_{\Sigma_t}{\rm d}^3x \,\pi^{ij}(x)\gamma_{ij}(x)
\ee
is
\be\nonumber
\{T,P\}=-1.
\ee
So, they form a global canonical pair.

Then, we extract the zero mode of the field $\pi(x)$, connected with the fields $N$ and ${\bf N}$
\be\nonumber
\pi=\frac{2}{N}\left(\sqrt\gamma\nabla_i N^i-\frac{{\rm d}}{{\rm d}t}\sqrt\gamma\right),
\ee
present it as a sum
\be\label{extractpi}
\pi (x)=\sqrt\gamma (x)<\pi>+\bar{\pi}(x),
\ee
where the mean value of $\pi(x)$ over a hypersurface $\Sigma_t$ is
\be\label{meanpi}
<\pi>:=\frac{\int_{\Sigma_t}\,{\rm d}^3 y \pi (y)}{\int_{\Sigma_t}\,
{\rm d}^3 y \sqrt{\gamma} (y)}.
\ee

The second term in (\ref{extractpi}) is the residue of the $\pi (x)$ with
zero mean value over a hypersurface $\Sigma_t$
\be\label{residuepi}
\int\limits_{\Sigma_t}\, {\rm d}^3 x\bar\pi (x)=0.
\ee
Thus, the mapping of the phase space $\Gamma_D$ onto the phase space $\bar\Gamma$ after
extraction of the global variables $T$ and $H$ is executed:
\be\nonumber
(D,\pi ;\tilde\gamma_{ij},\tilde\pi^{ij})\mapsto
(T, P; \bar{D},\bar\pi; \tilde\gamma_{ij},\tilde\pi^{ij}).
\ee
New variables are expressed via the old ones by the following formulae:
\bea
T(t)&=&-
\frac{\int_{\Sigma_t}{\rm d}^3y\,\sqrt{f}(y)D(y)\exp [(3/2)D(y)]}
{\int_{\Sigma_t}{\rm d}^3y\,\sqrt{f}(y)\exp [(3/2)D(y)]},
\nonumber \\
P(t)&=&\int_{\Sigma_t}{\rm d}^3y\,\pi (y),
\nonumber \\
\overline{D}(x)&=&D(x)-
\frac{\int_{\Sigma_t}{\rm d}^3y\,\sqrt{f}(y)D(y)\exp [(3/2)D(y)]}
{\int_{\Sigma_t}{\rm d}^3y\,\sqrt{f}(y)\exp [(3/2)D(y)]},
\nonumber\\
\overline{\pi}(x)&=&\pi(x)-\sqrt{f}(x)\exp [(3/2)D(x)]
\nonumber\\ &\times&
\frac{\int_{\Sigma_t}{\rm d}^3y\,\pi (y)}
{\int_{\Sigma_t}{\rm d}^3y\,\sqrt{f}(y)\exp [(3/2)D(y)]}. \nonumber
\eea
The Poisson brackets of the variables $(T, P, \bar{D}, \bar\pi)$ appear to be nonlinear, they read
\bea \nonumber
&& \{T,P\}=-1, \qquad
\{\bar{D}(x),P\}=0,
\\ \label{PoissonPiT}
&& \{\bar\pi (x),P\}=0, \qquad
\{\bar{\pi}(x), T\}=\frac{3}{2V_t}\sqrt{\gamma}(x)\bar{D}(x),
\label{PoissonTf} \\
&& \{T,\bar{D}(x)\}=0,\qquad
\{\bar\pi (x), \bar{D}(y)\}=-\frac{1}{V_t}\sqrt{\gamma}(x). \label{PoissonPif}
\eea

%%%%%%%%%%%%%%%%%%%%%%%%%%%%%%%%%%%%%%%%%%%%%%%%%%%%%%%%%%%%%%%%%%%%%%%%%%%%%%%%%%%%%%%%%%%%%%%%%
\section{Hamiltonian reduction and deparametrization} \label{Sect:reduction}
%%%%%%%%%%%%%%%%%%%%%%%%%%%%%%%%%%%%%%%%%%%%%%%%%%%%%%%%%%%%%%%%%%%%%%%%%%%%%%%%%%%%%%%%%%%%%%%%%

The integral Hamiltonian reduction from $\bar\Gamma$ to the phase space $\widetilde\Gamma_D$ with co-dimension 2 and
the parametrization procedure can be performed now
\be \nonumber
(T, P; \bar{D},\bar\pi; \tilde\gamma_{ij},\tilde\pi^{ij})
\mapsto (\bar{D},\bar\pi;\tilde\gamma_{ij},\tilde\pi^{ij}).
\ee
Thus, the reduced phase space of the cotangent bundle $T^{*}{\rm WDW}$ will be obtained.
To do that, we take $\pi(x)=2\sqrt\gamma K$ expressed from the Hamiltonian
constraint~(\ref{H}) as the Hamiltonian density, and let $T(t)$~(\ref{global}) to play
the role of the global time of the system (universe).
The reduced Poisson structure consists of (\ref{PoissonG}), (\ref{PoissonPi}), and (\ref{PoissonPif}).
The symplectic potential $\omega^1$~(\ref{symplectic}) reads
\bea \label{symplecticD}
\omega^1 &=& \tilde\pi^{ij}{\rm d}\tilde\gamma_{ij}+\pi{\rm d}{D}=
\tilde\pi^{ij}{\rm d}\tilde\gamma_{ij}+\bar\pi{\rm d}\bar{D}
\nonumber \\
&+& \sqrt\gamma <\pi>{\rm d}\bar{D}-
\pi{\rm d}T.
\eea
The third interference term of (\ref{symplecticD}) is exact differential form because the integral over the hypersurface
\bea \nonumber
&& \int\limits_{{t}_I}^{{t}_0}\,{\rm d}t<\pi>\int\limits_{\Sigma_t} {\rm d}^3x\sqrt\gamma (x)
\frac{{\rm d}\bar{D}}{{\rm d}t}=
\int\limits_{{t}_I}^{{t}_0}\,{\rm d}t<\pi><\frac{{\rm d}\bar{D}}{{\rm d}t}>
\\ \nonumber && \qquad
\equiv F(t_I,t_0)
\eea
is explicitly calculated.
Thus we obtain the ADM reduced action
\bea \label{Sreduced}
S &=& \int\limits_{T_I}^{T_0} {\rm d}T\left[\int\limits_{\Sigma_t} {\rm d}^3x
\left(\tilde\pi^{ij}\frac{{\rm d}\tilde\gamma_{ij}}{{\rm d} T}
+\bar\pi\frac{{\rm d}\bar{D}}{{\rm d}T}\right)- H\right]
\nonumber \\
&-& \int\limits_{{\rm T}_I}^{{\rm T}_0}\,
{\rm d}T \frac{{\rm d}t}{{\rm d}T}\int\limits_{\Sigma_t} {\rm d}^3x\, N^i{\mathcal H}_i.
\eea

The Hamiltonian depends on the global time $T$ explicitly
\be\label{reducedHam}
H[T(t),\bar{D}({x});\tilde\pi^{ij}(x),\tilde\gamma_{ij}(x)]:=
\int\limits_{\Sigma_t}\,{\rm d}^3x\,{\mathcal H} (x)
\ee
with the Hamiltonian density
\be\label{Hamdensity}
 {\mathcal H} (x)=4\sqrt{3}\sqrt{f}
 \biggl[\frac{1}{8\phi^{12}}\tilde\pi_{ij}\tilde\pi^{ij}
   + \frac{1}{\phi^{5}}\left(\tilde\Delta-\frac{1}{8}\tilde{R}\right)\phi
   + \frac{1}{8}\tilde{T}_{\bot\bot}\biggr]^{1/2}.
\ee
Functions of the field $\phi (x)$ in Eq.~(\ref{Hamdensity}) can be expressed
via the global time and the residues, {\it e.g.}, function $\phi^{6} (x)$ is a ratio of two parts:
\be\label{phi6}
\phi^6(x)=\sqrt{\frac{\gamma}{f}}(x)= \left(\frac{e^{\bar{D}(x)}}{e^{T(t)}}\right)^{3/2}.
\ee
The Hamiltonian density (\ref{Hamdensity}) can be rewritten in the explicit form
\bea
{\mathcal H} (x) &=& 4\sqrt{3}\sqrt{f}
\biggl[\frac{1}{8}\left(\frac{e^{\bar{D}}}{e^T}\right)^3\tilde\pi_{ij}\tilde\pi^{ij}
+ \left(\frac{e^T}{e^{\bar{D}}}\right)^{5/4}
\tilde\Delta \left(\frac{e^{\bar{D}}}{e^T}\right)^{1/4}
\nonumber \\
&-&\frac{1}{8}\left(\frac{e^T}{e^{\bar{D}}}\right)\tilde{R}
+\frac{1}{8}\tilde{T}_{\bot\bot}\biggr]^{1/2}.\label{Hamdensityexplicit}
\eea

The residue field $\bar{D}$ does not depend on the global time $T$:
\be
\frac{{\rm d}\bar{D}}{{\rm d}T}=0
\ee
on the Hamiltonian equations of motion.
It is not dynamical field.
Hence, the scale function (\ref{Psifactor}) is factorized as a product of two
functions depending, correspondingly, on time and space:
\be\nonumber
\phi^4 (x)=\exp(-T(t))\exp(\bar{D}({\bf x})).
\ee
The Hamiltonian equation of motion for the field $\bar\pi (x)$ is obtained by the bracket (\ref{PoissonPif})
\bea
\frac{{\rm d}\bar{\pi}}{{\rm d}T} &=& \int\limits_{\Sigma_t}\, {\rm d}^3y\,\{\bar{\pi} (x),\bar{D}(y)\}
\frac{\delta {H}}{\delta\bar{D}(y)}
\nonumber \\
&=& -\frac{\sqrt\gamma (x)}{V_t}\int\limits_{\Sigma_t}\, {\rm d}^3y\,
\frac{\delta {H}}{\delta\bar{D}(y)}=-\sqrt\gamma (x) F(T),
\eea
where $F(T)$ is the function of the time $T$.

The symplectic potential $\omega^1$~(\ref{symplectic}) in the extrinsic time approach was transformed with use
\bea \label{symplecticYork}
\pi\frac{{\rm d}D}{{\rm d}t} &=& 2K\sqrt\gamma\frac{{\rm d}D}{{\rm d}t}
=\frac{4}{3}K\frac{{\rm d}\sqrt\gamma}{{\rm d}t}
=\frac{{\rm d}}{{\rm d}t}\left(\frac{4}{3}K\sqrt\gamma\right)
\nonumber \\
&-& \sqrt\gamma\frac{{\rm d}}{{\rm d}t}\left(\frac{4}{3}K\right).
\eea
Four thirds of the extrinsic curvature scalar is the York time~\cite{York:1972sj}
\be\nonumber
\tau :=\frac{4}{3}K=\frac{2}{3}\frac{\pi}{\sqrt\gamma}.
\ee
Imposing the constant mean curvature gauge: the York time interval is equal to the coordinate time one $\tau=t$ sets the foliation of the spacetime.
From the local commutation relation of the fields
\be\nonumber
\{\sqrt\gamma (x),\tau (y)\}=\delta (x-y)
\ee
one obtains the global relation
\be\nonumber
\{V_t,T\}=1,
\ee
where the Hamiltonian is the volume of the hypersurface $V_t=\int\limits_{\Sigma_t}\,{\rm d}^3y\,\sqrt\gamma (y)$
and the global York time is defined as the integrand $T=\int\limits_{\Sigma_t}\,{\rm d}^3y\,\tau (y)$.

Our approach is different because the essence of these geometric
characteristics is treated in the opposite way. Note that in our approach the integral
characteristics $P(t)$ of the field $\pi(x)$ is the canonical partner of the global time.
In general, canonical momenta can't be defined within a hypersurface.
They should refer to the motion in time of the original $\Sigma_t$.
While the intrinsic time is the variable constructed entirely out of the metric
of the hypersurface. In this way, we establish the roles of the Hamiltonian and the
global time in accord with the general principles.

%%%%%%%%%%%%%%%%%%%%%%%%%%%%%%%%%%%%%%%%%%%%%%%%%%%%%%%%%%%%%%%%%%%%%%%%%%%%%%%%%%%%%%%%%%%%%%%%%
\section{Residue terms in a perturbed Friedmann universe} \label{Sect:residue}
%%%%%%%%%%%%%%%%%%%%%%%%%%%%%%%%%%%%%%%%%%%%%%%%%%%%%%%%%%%%%%%%%%%%%%%%%%%%%%%%%%%%%%%%%%%%%%%%%

Let us consider spatial metric perturbations in a Friedmann model and elucidate the role of
the residue term in~(\ref{extract}). For this purpose we use the harmonic analysis
of linear geometric perturbations using irreducible representations of the isometry group
of a constant curvature space. The eigenfunctions of the Laplace--Beltrami operator form
a basis of unitary representations of the group~\cite{Fock}
\bea \label{eigenfunctionssphere}
Y_{n l m} (\chi, \theta, \varphi) &=& 2^l l!
\sqrt{\frac{2(n+1)(n-l)!}{\pi(n+l+1)!}}\sin^l\chi
\nonumber \\ &\times&
C_{n-l}^{l+1}(\cos \chi)Y_{lm}(\theta, \varphi).
\eea
Here, $C_{n-l}^{l+1}(\cos \chi)$ are the Gegenbauer polynomials~\cite{Ryzhik}, and
$Y_{lm}(\theta, \varphi)$ are spherical harmonics. Indices run over the following values
\be \nonumber
n=0,1,2,\ldots;\quad l=0,1,\ldots,n; \quad m=-l,-l+1,\ldots, l.
\ee
The first eigenfunctions are
\bea\label{Yzero}
&& Y_{000}=\frac{1}{\sqrt{2}\pi},\quad Y_{100}=\frac{\sqrt{2}}{\pi}\cos\chi,
\\ &&
Y_{110}=\frac{\sqrt{2}}{\pi}\sin\chi\cos\theta,
\qquad
Y_{11,\pm 1}=\pm\frac{1}{\pi}\sin\chi\sin\theta e^{\pm\imath\varphi}.\nonumber
\eea
The orthogonality condition and normalization of the basis functions~(\ref{eigenfunctionssphere}) are well known
\bea
&& \int\limits_0^\pi d\chi\sin^2\chi\int\limits_0^\pi
d\theta\sin\theta\int\limits_0^{2\pi} d\varphi\,Y_{n l m}^{*}(\chi, \theta, \varphi)
Y_{n' l' m'}(\chi, \theta, \varphi)
\nonumber \\ && \qquad
=\delta_{n n'}\delta_{ll'}\delta_{mm'}.\label{ort}
\eea
Omitting the quantum indexes of the basis functions~(\ref{eigenfunctionssphere}),
the equation for eigenvalues can be represented in the following symbolic form
\begin{equation}\label{LaplaceB}
(({}^1\bar{\rm\Delta})+k^2)Y^{(s)}=0,
\end{equation}
where $-k^2$ is an eigenvalue of the Laplace--Beltrami operator $({}^1\bar{\rm\Delta})$
on a 3-sphere of unit radius. The connection $({}^1\bar\nabla)$ is associated with
this metric $({}^1 f_{ij})$.
For a positive curvature space we have $k^2=n(n+2)$.
So, we get the expressions for the residue terms in the countable basis
$Y_{nlm}(\chi,\theta,\varphi)$ of the Hilbert functional space
\bea \nonumber
&& \bar{D}(x)=
\sum\limits_{n=1}^\infty\sum\limits_{l=0}^n\sum\limits_{m=-l}^l \gamma_{nlm}Y_{nlm}(\chi,\theta,\varphi),
\\  \nonumber
&& \overline{\pi}(x) = \sum\limits_{n=1}^\infty\sum\limits_{l=0}^n\sum\limits_{m=-l}^l
\pi_{nlm}Y_{nlm}(\chi,\theta,\varphi)
\eea
with the expansion coefficients $\gamma_{nlm}$ and $\pi_{nlm}$.

%%%%%%%%%%%%%%%%%%%%%%%%%%%%%%%%%%%%%%%%%%%%%%%%%%%%%%%%%%%%%%%%%%%%%%%%%%%%%%%%%%%%%%%
\section{Hamiltonian equations of motion} \label{Sect:eq}
%%%%%%%%%%%%%%%%%%%%%%%%%%%%%%%%%%%%%%%%%%%%%%%%%%%%%%%%%%%%%%%%%%%%%%%%%%%%%%%%%%%%%%%

The momentum constraints generate spatial diffeomorphisms.
Below, by studying the dynamics, we choose the semi-geodesic slicing,
ignoring the action of the generators of diffeomorphisms~(\ref{constraintMom}).
We yield the Hamiltonian system without constraints on the contact manifold
$\mathbb{R}\times T^{*}{\rm WDW}$ with the intrinsic global time $T$.
The energy of the universe in our situation is not conserved,
it exponentially increases in time $T$.
The Hamiltonian flow is governed by the Hamiltonian~(\ref{reducedHam})
on the quadratic Lie---Poisson algebra of the generators $(\tilde\gamma_{ij}, \tilde\pi^{ij})$
(\ref{PoissonG}), (\ref{PoissonPi}) (see details in Appendix):
\bea \label{dgamma}
&& \frac{{\rm d}}{{\rm d}T}{\tilde\gamma}_{ij}({x})
= \int\limits_{\Sigma_t}\,{\rm d}^3{x}'
\{\tilde{\gamma}_{ij}({x}),{\tilde\pi}^{kl}({x}')\}
\frac{\delta}{\delta\tilde{\pi}^{kl}({x}')} H,
\\ \label{dpi}
&& \frac{{\rm d}}{{\rm d}T}{\tilde\pi}^{ij}({x})
= \int\limits_{\Sigma_t}\,{\rm d}^3{x}'
\{\tilde{\pi}^{ij}({x}),{\tilde\pi}^{kl}({x}')\}
\frac{\delta}{\delta\tilde{\pi}^{kl}({x}')} H
\nonumber \\ && \qquad
+\int\limits_{\Sigma_t}\,{\rm d}^3{x}'
\{\tilde{\pi}^{ij}({x}),{\tilde\gamma}_{kl}({x}')\}
\frac{\delta}{\delta\tilde{\gamma}_{kl}({x}')} H.
\eea
The functional derivative with respect to the momentum density reads
\be\label{varHpi}
\frac{\delta}{\delta\tilde{\pi}^{kl}({x}')} H [\phi;\tilde\pi^{ij},\tilde\gamma_{ij}]
=\frac{6\gamma (x')}{\phi^{12}(x'){\mathcal H}[\phi;\tilde\pi^{ij},\tilde\gamma_{ij};x')}
\tilde\pi_{kl}(x').
\ee
The derivative of the conformal metric with respect to the global time~(\ref{dgamma})
after application of~(\ref{PoissonG}) and~(\ref{varHpi}) becomes
\be \label{dgammadT}
\frac{{\rm d}}{{\rm d}T}\tilde\gamma_{ij}(x)
= \frac{12\gamma (x)\tilde\pi_{ij}(x)}
{\phi^{12}(x){\mathcal H}[\phi; \tilde\pi^{ij},\tilde\gamma_{ij};x)}.
\ee
Thus, the relation between the derivative of the generalized coordinates
$\tilde\gamma_{ij}$ with respect to the global time $T$ with the conjugate
generalized momenta $\tilde\pi^{ij}$ is obtained.

The first term in~(\ref{dpi}) is calculated easily
\bea \label{pipi}
&& \int\limits_{\Sigma_t}\,{\rm d}^3{x}'
\{\tilde{\pi}^{ij}({x}),{\tilde\pi}^{kl}({x}')\}
\frac{\delta}{\delta\tilde{\pi}^{kl}({x}')} H
\nonumber \\ && \qquad
= -\frac{2\gamma (x)\tilde\gamma^{ij}(x)\tilde\pi^{kl}(x)\tilde\pi_{kl}(x)}
{\phi^{12}(x){\mathcal H}[\phi; \tilde\pi^{ij},\tilde\gamma_{ij};x)}.
\eea

One can write the functional derivative of the Hamiltonian density $\mathcal{H}$
with respect to the conformal metric components $\tilde\gamma_{kl}$ as
\bea \nonumber
&& \frac{\delta}{\delta\tilde\gamma_{kl}(x)}{H}[\phi;\tilde\gamma_{ij},\tilde\pi^{ij}]
= \int\limits_{\Sigma_t}\,{\rm d}^3{y}\biggl(
-\frac{3\gamma(y)}{\phi^4(y){\mathcal H}(y)}
\frac{\delta}{\delta\tilde\gamma_{kl}(x)}
\\ &&
\times \tilde{R}[\tilde\gamma_{ij},y)
+ \frac{24\gamma(y)}{\phi^5(y){\mathcal H}(y)}
\frac{\delta}{\delta\tilde\gamma_{kl}(x)}\tilde\Delta_{y}\phi(y)\biggr). \nonumber
\eea
The functional derivative of the Ricci scalar with respect to the metric
coefficients reads
\bea \nonumber
&& \frac{\delta}{\delta\tilde\gamma_{kl}(x)}R[\tilde\gamma_{ij};y)=
\biggl(
-\tilde{R}^{kl}[\tilde\gamma_{kl};y)+\tilde\gamma^{kl}(y)
\tilde\Delta_{y}
- \tilde\nabla_{y}^k\tilde\nabla_{y}^l\biggr)
\\ \nonumber && \qquad \times
\delta(x-y).
\eea
By summing up these four terms we obtain the functional derivative
\bea \label{foursum}
&& \frac{\delta}{\delta\tilde\gamma_{kl}(x)}H[\phi,\tilde\gamma_{ij},\tilde\pi^{ij}]
\frac{3\gamma (x)}{\phi^4(x){\mathcal H}(x)}\tilde{R}^{kl}
\\ \nonumber && \qquad
+ 3\left(\tilde\nabla^k_x\tilde\nabla^l_x-\tilde\gamma^{kl}(x)\tilde\Delta_x\right)
\left(\frac{\gamma (x)}{\phi^4(x){\mathcal H}(x)}\right)\nonumber
\\ \nonumber && \qquad
+ 12(2\tilde\gamma^{km}\tilde\gamma^{ln}-\tilde\gamma^{kl}\tilde\gamma^{mn})\tilde\nabla_m
\left(\frac{\gamma (x)}{\phi^5(x){\mathcal H}(x)}\right)\tilde\nabla_n\phi (x).
\eea

Finally, the derivative of the conformal momentum density with respect to
the global time~(\ref{dpi}) with application of the commutation relations between
conformal phase variables and taking into account~(\ref{pipi}) and~(\ref{foursum}),
becomes
\bea \label{dpidT}
&& \frac{\rm d}{{\rm d}T}\tilde\pi^{ij}(x)
= - \frac{6\gamma (x)}{\phi^4(x){\mathcal H}(x)}\left(\tilde{R}^{ij}
- \frac{1}{6}\tilde\gamma^{ij}\tilde{R}\right)
\\ \nonumber && \quad
- \frac{2\gamma (x)}{\phi^{12}(x){\mathcal H}(x)}\tilde\gamma^{ij}(x)
\tilde\pi^{kl}\tilde\pi_{kl}
\\ \nonumber && \quad
- 3\biggl(\tilde\nabla^i\tilde\nabla^j
+ \tilde\nabla^j\tilde\nabla^i-\frac{4}{3}\tilde\gamma^{ij}
\tilde\nabla^k\tilde\nabla_k\biggr)
\left[\frac{\gamma (x)}{\phi^4 (x){\mathcal H}(x)}\right]
\\ \nonumber && \quad
- 24\left(\tilde\gamma^{ik}\tilde\gamma^{jl}
+ \tilde\gamma^{jk}\tilde\gamma^{il}
- \frac{2}{3}\tilde\gamma^{ij}\tilde\gamma^{kl}\right)\tilde\nabla_l\phi\tilde\nabla_k
\left[\frac{\gamma (x)}{\phi^5 (x){\mathcal H}(x)}\right].
\eea

%%%%%%%%%%%%%%%%%%%%%%%%%%%%%%%%%%%%%%%%%%%%%%%%%%%%%%%%%%%%%%%%%%%%%%%%%%%%%%%%%%%%%%%%%%%%%%%%%
\section{Relations between time intervals} \label{Sect:times}
%%%%%%%%%%%%%%%%%%%%%%%%%%%%%%%%%%%%%%%%%%%%%%%%%%%%%%%%%%%%%%%%%%%%%%%%%%%%%%%%%%%%%%%%%%%%%%%%%

To obtain the relation between an intrinsic time interval and a coordinate time
interval one should find the derivative of a hypersurface volume with respect
to the coordinate time
\be\label{Plateau}
\frac{{\rm d}V_t}{{\rm d}t}=\int\limits_{\Sigma_t}{\rm d}^3x
\frac{\partial}{\partial t}\sqrt\gamma
= \int\limits_{\Sigma_t}{\rm d}^3x\left(-NK+\nabla_i N^i\right)\sqrt\gamma.
\ee
The above equality follows from the definition~(\ref{Kijdef})
\be \nonumber
\nabla\cdot{\bf N}-NK=\frac{1}{2}\gamma^{ij}\frac{{\rm d}\gamma_{ij}}{{\rm d}t}
=\frac{1}{\sqrt\gamma}\frac{\rm d}{{\rm d}t}\sqrt\gamma.
\ee
The second integrand is zero according to the Gauss--Ostrogradsky theorem
\be \nonumber
\int\limits_{\Sigma_t}{\rm d}^3x\,\nabla_i N^i\sqrt\gamma=\int\limits_{\Sigma_t}{\rm d}^3x\,
\frac{\partial}{\partial x^i}\left(N^i\sqrt\gamma\right)=0.
\ee
Hence, the speed of the volume change depends on the lapse function only
\be \label{volumechange}
\frac{{\rm d}V_t}{{\rm d}t}=-\int\limits_{\Sigma_t}{\rm d}^3x\sqrt\gamma\,N K.
\ee
The maximal slicing corresponds to the vanishing of the scalar $K=0$.
Analogously, the relation between the intrinsic time interval ${\rm d}D$ (\ref{DiracTpi})
and the coordinate time interval ${\rm d}t$ is obtained
\be\label{dDdt}
\frac{{\rm d}D}{{\rm d}t}=\frac{2}{3}\left(NK-\nabla_i N^i\right).
\ee
It is composed of normal and tangent components.

In this way, we obtain the relation between the intrinsic global time interval
${\rm d}T$ (\ref{global}) and the coordinate time interval ${\rm d}t$
\be\label{dTdt}
\frac{{\rm d}T}{{\rm d}t}=
-<\frac{{\rm d}D}{{\rm d}t}>-<D\frac{{\rm d}\sqrt\gamma}{{\rm d}t}>+<D>\frac{{\rm d}V_t}{{\rm d}t}.
\ee
The derivatives in (\ref{dTdt}) are defined in (\ref{Plateau}), (\ref{volumechange}), and (\ref{dDdt}).
The conformal Friedmann equation also can be written if the conformal time interval
is introduced $d\eta=dt/a$. The lapse function $N$ and the shift vector ${\bf N}$
are contained in (\ref{dTdt}). According to the thin sandwich conjecture~\cite{Bartnik:1993gh},
they can be found from the theory.
In the case of semi-geodesic slicing ${\bf N}=0$, one obtains the simple relation
between the time intervals $d\tau= N dt$,
demonstrating the notion of the lapse of the proper watch.

The application of the perfect cosmological principle
to the Friedmann interval yields the ratio of universe radius
$a(t)$ to the present day one $a_0$ from the standard cosmological conception
\be\nonumber
\left(\frac{a(t)}{a_0}\right)^2=e^{-T}.
\ee
Here, for the background space a sphere of the present day radius $a_0$ is taken.
The global time is related to the observable redshift~\cite{Weinberg} in the standard way:
\be \nonumber
z(t)=\frac{a_0-a(t)}{a(t)}.
\ee
The cosmological singularity reached under $t=0$ corresponds to $T\to +\infty$. The concept of transition from finite time interval to the infinite one is in accordance with the infinite logarithmic time of life after the Big Bang including infinite chains of events \cite{Belinski}.

%%%%%%%%%%%%%%%%%%%%%%%%%%%%%%%%%%%%%%%%%%%%%%%%%%%%%%%%%%%%%%%%%%%%%%%%%%%%%%%%%%%%%%%%%%%%%%%%%
\section{Discussion and conclusions} \label{Sect:concl}
%%%%%%%%%%%%%%%%%%%%%%%%%%%%%%%%%%%%%%%%%%%%%%%%%%%%%%%%%%%%%%%%%%%%%%%%%%%%%%%%%%%%%%%%%%%%%%%%%

We demonstrated that in Geometrodynamics of closed manifolds,
it is possible to generalize the Misner approach and introduce the global time.
After application of the Hamiltonian reduction procedure, we got differential
evolution equations for conformal metric components and conformal momentum densities.
They do not contain Lagrange multipliers contrary to the ones obtained in~\cite{Arnowitt}.
In opposite to the case of asymptotically flat spacetimes, for closed manifolds we got
a non-conservative Hamiltonian systems.

It was assumed~\cite{Isenberg:1981fa,Fischer:1996qg,FischerRicci,Fischer2} that
the extrinsic time is the most useful variable in dealing with Einstein solutions
on spatially compact surfaces.
Our approach is different since these geometric characteristics possess in our case quite
an opposite essence. In general, we claim that canonical momenta are not defined within
the hypersurface. They refer to the motion in time of the original $\Sigma_t$.
And the intrinsic time is the variable constructed entirely out of the metric of
the hypersurface. So, the roles of the Hamiltonian and global time are interchanged
with respect to the ones adopted in
refs.~\cite{Isenberg:1981fa,Fischer:1996qg,FischerRicci,Fischer2,Valentini}.
As discussed in Sect.~\ref{Sect:globaltime}, our treatment of the Hamiltonian and the
global time is in accord with the general principles.

The deviation from the mean value of the global time behaves as a classical scalar non-dynamical field.
It deserves an additional attention to be physically interpreted. Note that it emerged
without any modification of the Einstein's theory. The Wheeler's thin sandwich conjecture
in General Relativity known to be valid even for higher dimensional theories of gravity
under positive lapse and some restriction~\cite{Avalos}.
Thus, the lapse function can be found from the Hamiltonian constraint.
It was conjectured~\cite{Arbuzov:2010fz}, that the deviations from the mean value
of the global time can play the role of static gravitational potentials.

In the extrinsic time approach, the program of the Hamiltonian reduction carried out under the condition that
the spacetime admits the constant mean curvature foliation. In opposite, we do not use any gauge yet.
The reduced Einstein's flow was described by a time-dependent non-local reduced Hamiltonian.
In our case, the Hamiltonian of the problem is presented in the explicit form that allows to write the Hamiltonian
flow.

Earlier to construct a scalar, the Minkowskian metric as a background one was used
for asymptotically flat spaces~\cite{Gor}. The intrinsic time interval $\delta D$ as
a scalar field was implemented in the symplectic 1-form~\cite{Murchadha:2012zz,Ita:2017fwh}.
For splitting one degree of freedom, the average of the trace of the momentum density
was used as the York time in the shape dynamics~\cite{Mercati:2014ama}. The key difference
of our study is the consideration of closed manifolds without the asymptotically flat space
condition. Our choice is motivated by cosmological applications where models with
closed manifolds are of great interest.

For interpretation of the latest data of the Hubble diagram, the global time
as the scale factor of the Friedmann model was successfully
implemented in refs.~\cite{Zakharov:2010nf,Pervushin:2017zfj,PavlovMIPh}.
The choice of conformal variables allows to suggest a new interpretation of
the redshift of distant stellar objects. Both the changing volume of the Universe
in standard cosmology and the changing of masses of elementary particles
in conformal cosmology~\cite{Narlikar} can serve as the measure of time.
We have shown that the cosmological observable redshift effect can be directly related
to the global time of the Universe. In general, we claim that the intrinsic time can
be considered as an evolution parameter of cosmological evolution.

%%%%%%%%%%%%%%%%%%%%%%%%%%%%%%%%%%%%%%%%%%%%%%%%%%%%%%%%%%%%%%%%%%%%%%%%%%%%%%%%%%%%%%%%%%%%%%%%%%%%
\begin{acknowledgments}
A.E.~Pavlov thanks to the Bogoliubov Laboratory of Theoretical Physics of
the Joint Institute for Nuclear Research (Dubna) for hospitality.
A.E.~Pavlov obliged to Prof. A.E.~Fischer for acquaintance with his papers
on Hamiltonian reduction of Einstein's equations and conformal Ricci flows
and thanks Prof. M.~Henneaux for presenting the book on modern algebraic description 
of the cosmological singularity.
\end{acknowledgments}
%%%%%%%%%%%%%%%%%%%%%%%%%%%%%%%%%%%%%%%%%%%%%%%%%%%%%%%%%%%%%%%%%%%%%%%%%%%%%%%%%%%%%%%%%%%%%%%%%
\section*{Appendix} \label{Sect:App}
%%%%%%%%%%%%%%%%%%%%%%%%%%%%%%%%%%%%%%%%%%%%%%%%%%%%%%%%%%%%%%%%%%%%%%%%%%%%%%%%%%%%%%%%%%%%%%%%%

To obtain functional derivatives we need some useful results for variations with respect to metric
in Banach functional space
\bea \nonumber
&& \delta\gamma=\gamma\gamma^{ij}\delta\gamma_{ij}=-\gamma\gamma_{ij}\delta\gamma^{ij},\qquad
\delta\sqrt\gamma=\frac{1}{2}\sqrt\gamma\gamma^{ij}\delta\gamma_{ij},
\\ \nonumber
&& \delta\gamma^{ij}=-\gamma^{ik}\gamma^{jl}\delta\gamma_{kl},\qquad
\frac{\delta}{\delta\gamma_{ij}(x)}V_t=\frac{1}{2}\sqrt\gamma (x)\gamma^{ij}(x).
\eea

$\bullet$
Let us prove the statement
\be\nonumber
\{T,P\}=-1.
\ee
Calculate the functional derivative of the ratio of two functionals (\ref{mean}) by the metric
\bea \label{funcgamma}
&& \frac{\delta}{\delta\gamma_{ij}(x)}<\ln\sqrt{\frac{\gamma}{f}}>
=\frac{1}{2V_t}\sqrt{\gamma}(x)\gamma^{ij}(x)
\nonumber \\ && \qquad \times
\left(\ln\sqrt{\frac{\gamma}{f}}(x)+1-<\ln\sqrt{\frac{\gamma}{f}}>\right).
\eea
The functional derivative of $P$ (\ref{Pi}) is the metric tensor
\be\label{funcpi}
\frac{\delta P}{\delta\pi^{ij}(x)}=\gamma_{ij}(x).
\ee
Then we calculate the Poisson bracket of the functionals (\ref{mean}) and (\ref{Pi})
with the use of the derivatives (\ref{funcgamma}) and (\ref{funcpi})
\be\nonumber
\left\{<\ln\sqrt{\frac{\gamma}{f}}[\gamma_{ij}]>,P[\gamma_{ij},\pi^{ij}]\right\}=\frac{3}{2}.
\ee
The obtained bracket can be rewritten as
\be
-\frac{2}{3}\left\{\ln<\sqrt{\frac{\gamma}{f}}[\gamma_{ij}]>,P[\gamma_{ij},\pi^{ij}]\right\}=\{T,P\}=-1.
\ee
Thus we prove the canonical commutation relation and get the global time (\ref{global}).
The global time (\ref{global}) is a scalar. If one consider diffeomorphisms
\be\nonumber
\delta\gamma_{ij}=\pounds_{\bf N}\gamma_{ij}\equiv\nabla_i N_j+\nabla_j N_i,
\ee
then, applying it to (\ref{funcgamma}), one obtains
\bea \nonumber
&& \delta<\ln\sqrt{\frac{\gamma}{f}}>=
\frac{1}{2V_t}\sqrt{\gamma}(x)\gamma^{ij}(x)(\nabla_i N_j+\nabla_j N_i)
\\ \nonumber && \qquad \times
\left(\ln\sqrt{\frac{\gamma}{f}}(x)+1-<\ln\sqrt{\frac{\gamma}{f}}>\right).
\eea
Integrating over the hypersurface $\Sigma_t$ one gets the integral of the divergent term
\bea \nonumber
&& \int\limits_{\Sigma_t}\,{\rm d}^3x\delta<\sqrt{\frac{\gamma}{f}}>=
\frac{1}{V_t}\int\limits_{\Sigma_t}\,{\rm d}^3x\frac{\partial}{\partial x^i}
\biggl[\gamma^{ij}(x)N_j
\\ \nonumber && \times
\biggl(\ln\sqrt{\frac{\gamma}{f}}(x)+1-<\ln\sqrt{\frac{\gamma}{f}}>\biggr)\biggr]=0.
\eea
So, the global time is a scalar, {\it i.e.}, it is conserved under spatial diffeomorphisms.

$\bullet$
Let us calculate the bracket
\be\nonumber
\{\bar\pi (x),P\}=\{\pi (x),\int\limits_{\Sigma_t}{\rm d}^3y\,\pi (y)\}-\{\sqrt\gamma (x)<\pi>,P\}.
\ee
The first term is zero, calculate the second one. We need the following functional derivatives
\bea\nonumber
&&\frac{\delta}{\delta\gamma_{ij}(x')}\left(\sqrt{\gamma} (x)<\pi>\right)
=\frac{1}{2}\sqrt{\gamma} (x)\gamma^{ij}(x)\delta (x-x')<\pi>
\\ && \qquad
+ \sqrt{\gamma} (x)\frac{\delta}{\delta\gamma_{ij}(x')}<\pi>, \label{pi1}
\eea
\be\label{pi2}
\frac{\delta}{\delta\gamma_{ij}(x')}<\pi>=\frac{1}{V_t}\pi^{ij}(x')
-\frac{1}{2V_t}\sqrt{\gamma}(x')\gamma^{ij}(x')<\pi>,
\ee
\be\label{pi3}
\frac{\delta}{\delta\pi^{ij}(x')}P=\gamma_{ij}(x'),\qquad
\frac{\delta}{\gamma_{ij}(x')}P=\pi^{ij}(x'),
\ee
\be\label{pi4}
\frac{\delta}{\delta\pi^{ij}(x')}\left(\sqrt{\gamma} (x)<\pi>\right)=\frac{\sqrt\gamma (x)}{V_t}\gamma_{ij}(x').
\ee
Collecting all Eqs.~(\ref{pi1})--(\ref{pi4}), we obtain
\bea
&& \{\bar\pi (x),P\} = \frac{3}{2}\sqrt\gamma (x)<\pi>+\sqrt\gamma (x)<\pi>
\nonumber \\ && \qquad
- \frac{3}{2}\sqrt\gamma (x)<\pi>-\sqrt\gamma (x)<\pi>=0.
\eea

$\bullet$
Let us calculate the bracket
\be\nonumber
\{\bar\pi (x), \bar{D}(y)\}=\{\bar\pi (x), \bar{D}(y)\}-
\{\bar\pi (x), <D>\}.
\ee
Here we need the following functional derivatives
\bea\label{barpiij}
&& \frac{\delta}{\delta\pi^{ij}(x')}\bar\pi (x) = \gamma_{ij}(x)\delta (x-x')
- \frac{1}{V_t}\sqrt\gamma (x)\gamma_{ij}(x'),
\nonumber \\
&& \frac{\delta}{\delta\gamma_{ij}(x')}\bar{D}(y)
= \frac{1}{3}(y)\gamma^{ij}(y)\delta(y-x')
\\ \nonumber && \quad
- \frac{\sqrt\gamma (x')\gamma^{ij}(x')}{3V_t}
\biggl(\ln\sqrt{\frac{\gamma}{f}}(x')
+1-<\ln\sqrt{\frac{\gamma}{f}}>\biggr).
\eea \label{bargamma}
We get the bracket
\be
\{\bar\pi (x), \bar{D}(y)\}=
-\frac{1}{V_t}\sqrt\gamma (x).
\ee

$\bullet$
Variations for components of metric connections
\bea \label{deltaGamma}
&& \delta\Gamma^k_{ij}=\frac{1}{2}\gamma^{kl}\left(\nabla_i(\delta\gamma_{lj})
+\nabla_j(\delta\gamma_{li})
-\nabla_l(\delta\gamma_{ij})\right),
\nonumber \\
&& \delta\Gamma^k_{ik}=\frac{1}{2}\gamma^{kl}\nabla_i(\delta\gamma_{kl});
\eea
variations of components of Riemannian tensor
\be \nonumber
\delta R^i_{jkl}=\nabla_l\left(\delta\Gamma^i_{jk}\right)
-\nabla_k\left(\delta\Gamma^i_{jl}\right);
\ee
variations of components of Ricci tensor known as Palatini identity
\be \label{deltaRicci}
\delta R_{ij}=\nabla_j\left(\delta\Gamma^k_{ik}\right)-\nabla_k\left(\delta\Gamma^k_{ij}\right);
\ee
a variation of the Ricci scalar
\be \label{deltaRicciscalar}
\delta R=\delta (R_{ij}\gamma^{ij})
= -R^{ij}(\delta\gamma_{ij})+\gamma^{ij}\delta R_{ij}.
\ee
Using formulae (\ref{deltaGamma}) and (\ref{deltaRicci}), one gets
\bea \nonumber
&& \delta R_{ij}=\frac{1}{2}\nabla_j\nabla_i\left(\gamma^{kl}\delta\gamma_{kl}\right)
+ \frac{1}{2}\nabla_l\nabla^l(\delta\gamma_{ij})
\\ \nonumber && \qquad
- \frac{1}{2}\nabla_k\nabla_j\left(\gamma^{kl}\delta\gamma_{li}\right)-
\frac{1}{2}\nabla_k\nabla_i\left(\gamma^{kl}\delta\gamma_{lj}\right).
\eea
Contracting the tensor, we get
\bea \nonumber
\gamma^{ij}\delta R_{ij}=\gamma^{ij}\Delta\left(\delta\gamma_{ij}\right)-
\nabla^i\nabla^j\left(\delta\gamma_{ij}\right),
\eea
where
\bea \nonumber
\Delta:=\gamma^{ij}\nabla_i\nabla_j\equiv\frac{1}{\sqrt\gamma}\frac{\partial}{\partial x^i}\left(
\sqrt\gamma\gamma^{ij}\frac{\partial}{\partial x^j}\right)
\eea
is the invariant Laplace operator.

Finally, the expression for the Ricci scalar variation (\ref{deltaRicciscalar}) that we need are
\be
\delta{R}=-{R}^{ij}(\delta\gamma_{ij})+
\gamma^{ij}\Delta (\delta\gamma_{ij})-
\nabla^i\nabla^j(\delta\gamma_{ij}).
\ee

$\bullet$
Let us prove a useful formula. Here and below, using the Gauss--Ostrogradsky
theorem, we get rid of divergent terms:
\bea
&& \int\limits_{\Sigma_t}{\rm d}^3 x\, \sqrt\gamma (x) F(x)\Delta_x\delta (x-x')
=\int\limits_{\Sigma_t}{\rm d}^3 x\, F(x)
\nonumber \\ && \quad
\times \frac{\partial}{\partial x^i}\left(
\sqrt\gamma (x)\gamma^{ij}(x)\frac{\partial}{\partial x^j}\delta (x-x')\right)
\nonumber \\ && \quad
= - \int\limits_{\Sigma_t}{\rm d}^3 x\,\frac{\partial}{\partial x^i}(F(x))
\sqrt\gamma (x)\gamma^{ij}(x)\frac{\partial}{\partial x^j}\delta (x-x')
\nonumber \\ && \quad
= \int\limits_{\Sigma_t}{\rm d}^3 x\,\delta (x-x')\frac{\partial}{\partial x^j}
\left(\sqrt\gamma (x)\gamma^{ij}(x)
\frac{\partial}{\partial x^i}\left(F(x)\right)\right)
\nonumber \\ && \quad
= \int\limits_{\Sigma_t}{\rm d}^3 x\,\delta (x-x')\sqrt\gamma (x)\Delta_x F(x)=
\sqrt\gamma (x')\Delta_{x'}F(x').\nonumber
\eea
Thus we proved the Hermiticity of the Laplace operator.

We need also the relation
\bea
&& \int\limits_{\Sigma_t}{\rm d}^3 x\, \sqrt\gamma (x) F(x)\nabla^i_x\nabla^j_x\delta\gamma_{ij}
\nonumber \\ && \qquad
= \int\limits_{\Sigma_t}{\rm d}^3 x\, \sqrt\gamma (x) F(x)\gamma^{ki}\gamma^{lj}\nabla_k\nabla_l\delta\gamma_{ij}
\nonumber \\ && \qquad
= \int\limits_{\Sigma_t}{\rm d}^3 x\, \sqrt\gamma (x)\nabla_k\left(
F(x)\gamma^{ki}\gamma^{lj}\nabla_l\delta\gamma_{ij}\right)
\nonumber \\ && \qquad
- \int\limits_{\Sigma_t}{\rm d}^3 x\, \sqrt\gamma (x)\gamma^{ki}\gamma^{lj}\nabla_k F(x)
\nabla_l\delta\gamma_{ij}
\nonumber \\ && \qquad
- \int\limits_{\Sigma_t}{\rm d}^3 x\, \sqrt\gamma (x)\nabla_l\left(
\gamma^{ki}\gamma^{lj}\nabla_k F(x)\delta\gamma_{ij}\right)
\nonumber \\ && \qquad
+ \int\limits_{\Sigma_t}{\rm d}^3 x\,
\sqrt\gamma (x)\nabla^j\nabla^i F(x)\delta\gamma_{ij}. \nonumber
\eea
Taking a functional derivative of the expression by metric coefficients,
we prove the next useful formula for the Hermitian operator
\bea \nonumber
&& \int\limits_{\Sigma_t}{\rm d}^3 x\,\sqrt\gamma (x)F(x)\nabla_x^i\nabla_x^j\delta (x-x')
\nonumber \\ && \qquad
= \int\limits_{\Sigma_t}{\rm d}^3 x\,\sqrt\gamma (x)\delta (x-x')\nabla_x^i\nabla_x^j F(x).
\eea
Then, we simplify the following integral
\bea
&& \int\limits_{\Sigma_t}{\rm d}^3x\sqrt\gamma (x) F(x)\Delta_x\phi
\nonumber \\ && \qquad
= \int\limits_{\Sigma_t}{\rm d}^3x F(x)\frac{\partial}{\partial x^i}\left(
\sqrt\gamma\gamma^{ij}\frac{\partial\phi}{\partial x^j}\right)
\nonumber \\ && \quad
= \int\limits_{\Sigma_t}{\rm d}^3x \frac{\partial}{\partial x^i}\left(F(x)
\sqrt\gamma\gamma^{ij}\frac{\partial\phi}{\partial x^j}\right)
\nonumber \\ && \qquad
- \int\limits_{\Sigma_t}{\rm d}^3x\sqrt\gamma (x)\gamma^{ij}\frac{\partial F}{\partial x^i}
\frac{\partial\phi}{\partial x^j}.\nonumber
\eea
Now we can find a functional derivative of it
\bea
&& \frac{\delta}{\delta\gamma_{ij}(x')}\int\limits_{\Sigma_t}{\rm d}^3x\sqrt\gamma (x) F(x)\Delta_x\phi
\nonumber \\ && \qquad
= - \frac{1}{2}\int\limits_{\Sigma_t}{\rm d}^3x\sqrt\gamma (x)\gamma^{ij}\delta(x-x')\gamma^{kl}
\frac{\partial F}{\partial x^k}\frac{\partial\phi}{\partial x^l}
\nonumber \\ && \qquad
+\int\limits_{\Sigma_t}{\rm d}^3x\sqrt\gamma (x)\delta(x-x')\gamma^{ki}\gamma^{lj}
\frac{\partial F}{\partial x^k}\frac{\partial\phi}{\partial x^l}
\nonumber \\ && \qquad
= - \frac{1}{2}\int\limits_{\Sigma_t}{\rm d}^3x\sqrt\gamma (x)\gamma^{ij}
\delta(x-x')\Delta_1(F,\phi)
\nonumber \\ && \qquad
+ \int\limits_{\Sigma_t}{\rm d}^3x\sqrt\gamma (x)\delta(x-x')\gamma^{ki}\gamma^{lj}
\nabla_k F\nabla_l\phi
\nonumber \\ && \qquad
= - \frac{1}{2}\sqrt\gamma (x')\gamma^{ij}(x')\Delta_1 (F,\phi)
\nonumber \\ && \qquad
+ \sqrt\gamma (x')\gamma^{ki}(x')\gamma^{lj}(x')\nabla_k F\nabla_l\phi .\nonumber
\eea
Here, assuming $F$ and $\phi$ to be scalar fields,
we replaced the partial derivatives with covariant ones,
and introduced the Beltrami invariant operator
\be \nonumber
\Delta_1(F,\phi)=:\gamma^{kl}\nabla_k F\nabla_l \phi.
\ee

%%%%%%%%%%%%%%%%%%%%%%%%%%%%%%%%%%%%%%%%%%%%%%%%%%%%%%%%%%%%%%%%%%%%%%%%%%%%%%%%%%%%%%%%%%%%%%%%%%%%%%%%%%%

\end{document}